\documentclass[aps,twocolumn,amsmath,amssymb,floatfix,showpacs,
superscriptaddress]{revtex4}

\usepackage[dvips]{graphics}
\usepackage{color}
\definecolor{gold}{rgb}{0.85,0.66,0}
\definecolor{dblue}{rgb}{0,0,0.6}

\begin{document}

\title{\textcolor{dblue}{On the way to meet the experimental observation 
of persistent current in a mesoscopic cylinder: A mean field study}}

\author{Santanu K. Maiti}

\email{santanu.maiti@saha.ac.in}

\affiliation{Theoretical Condensed Matter Physics Division, Saha
Institute of Nuclear Physics, Sector-I, Block-AF, Bidhannagar,
Kolkata-700 064, India}

\affiliation{Department of Physics, Narasinha Dutt College, 129
Belilious Road, Howrah-711 101, India}

\begin{abstract}
The behavior of persistent current in a mesoscopic cylinder threaded by an
Aharonov-Bohm flux $\phi$ is carefully investigated within a Hartree-Fock
mean field approach. We examine the combined effect of second-neighbor
hopping integral and Hubbard correlation on the enhancement of persistent
current in presence of disorder. A significant change in current amplitude
is observed compared to the traditional nearest-neighbor hopping model
and the current amplitude becomes quite comparable to experimental
realizations. Our analysis is found to exhibit several interesting results
which have so far remained unaddressed.
\end{abstract}


\maketitle

\section{Introduction}

Over the last many years appearance of persistent current in metallic
single-channel rings and multi-channel cylinders has drawn much attention
in theoretical as well as in experimental research. In mesoscopic regime
where dimensions of a system are comparable to mean free path of an electron
the phase coherence of electronic states is of fundamental importance and 
the existence of dissipationless current in a mesoscopic conducting ring 
threaded by an Aharonov-Bohm (AB) flux $\phi$ is a direct consequence of 
quantum phase coherence. In this new quantum regime, two important 
aspects appear at low temperatures and they are as follows. \\
$\bullet$ The phase coherence length $L_{\phi}$ i.e., the length scale 
for which an electron maintains its phase memory, increases significantly 
with the lowering of temperature and becomes comparable to the system size 
$L$. \\
$\bullet$ The energy levels of these small finite size systems are 
discrete. \\
These two are the most essential criteria for the existence of persistent
current in a small metallic ring/cylinder due to the application of an 
external magnetic flux $\phi$. In the pioneering work of B\"{u}ttiker, 
Imry and Landauer~\cite{butt}, the appearance of persistent current in 
metallic rings has been explored. Later, many excellent 
experiments~\cite{levy,chand,mailly,jari,deb,reu} have been carried out 
in several ring and cylindrical geometries to reveal the actual mechanisms of 
persistent current. Though much efforts have been paid to study persistent
current both theoretically~\cite{cheu1,cheu2,peeters1,peeters2,peeters3,mont,mont1,alts,von,schm,ambe,abra,bouz,giam,yu,belu,ore,xiao1,xiao2,san1,san2,san3} 
as well as experimentally~\cite{levy,chand,mailly,jari,deb,reu}, yet 
several drawbacks still exist between the theory and experiment, and the 
full knowledge about it in this scale is not well established even today. 

The results of the single loop experiments are significantly different 
from those for the ensemble of isolated loops. Persistent currents with 
expected $\phi_0$ periodicity have been observed in isolated single Au 
rings~\cite{chand} and in a GaAs-AlGaAs 
ring~\cite{mailly}. Levy {\em et al.}~\cite{levy} found oscillations with 
period $\phi_0/2$ rather than $\phi_0$ in an ensemble of $10^7$ independent 
Cu rings. Similar $\phi_0/2$ oscillations were also reported for an ensemble 
of disconnected $10^5$ Ag rings~\cite{deb} as well as for an array of $10^5$ 
isolated GaAs-AlGaAs rings~\cite{reu}. In a recent experiment, Jariwala 
{\em et al.}~\cite{jari} obtained both $\phi_0$ and $\phi_0/2$ periodic 
persistent currents in an array of thirty diffusive mesoscopic Au rings. 
Except for the case of the nearly ballistic GaAs-AlGaAs ring~\cite{mailly}, 
all the measured currents are in general one or two orders of magnitude 
larger than those expected from the theory.

Free electron theory predicts that at $T=0$, an ordered one-dimensional 
metallic ring threaded by magnetic flux $\phi$ supports persistent current 
with maximum amplitude $I_0=ev_F/L$, where $v_F$ is the Fermi velocity 
and $L$ is the circumference of the ring. Metals are intrinsically 
disordered which tends to decrease the persistent current, and the 
calculations show that the disorder-averaged current $\langle I \rangle$
crucially depends on the choice of the ensemble~\cite{cheu2,mont,mont1}. 
The magnitude of the current $\langle I^2\rangle^{1/2}$ is however 
insensitive to the averaging issues, and is of the order of $I_0 l/L$, 
$l$ being the elastic mean free path of the electrons. This expression 
remains valid even if one takes into account the finite width of the ring
by adding contributions from the transverse channels, since disorder leads 
to a compensation between the channels~\cite{cheu2,mont}. However, the 
measurements on an ensemble of $10^7$ Cu rings~\cite{levy} reported a 
diamagnetic persistent current of average amplitude $3 \times 10^{-3}$
$ev_F/L$ with half a flux-quantum periodicity. Such $\phi_0/2$ 
oscillations with diamagnetic response were also found in other 
persistent current experiments consisting of ensemble of isolated
rings~\cite{deb,reu}.

Measurements on single isolated mesoscopic rings on the other hand 
detected $\phi_0$-periodic persistent currents with amplitudes 
of the order of $I_0 \sim ev_F/L$, (closed to the value for an ordered ring). 
Theory and experiment~\cite{mailly} seem to agree only when {\em disorder is
weak}. In another recent nice experiment Bluhm {\em et al.}~\cite{blu} have
measured the magnetic response of $33$ individual cold mesoscopic gold 
rings, one ring at a time, using a scanning SQUID technique. They have 
measured $h/e$ component and predicted that the measured current amplitude
agrees quite well with theory~\cite{cheu1} in a single ballistic 
ring~\cite{mailly} and an ensemble of $16$ nearly ballistic 
rings~\cite{raba}. However, the 
amplitudes of the currents in single-isolated-diffusive 
gold rings~\cite{chand} were two orders of magnitude larger than the 
theoretical estimates. This discrepancy initiated intense theoretical 
activity, and it is generally believed that the electron-electron 
correlation plays an important role in the disordered diffusive
rings~\cite{abra,bouz,giam}.
An explanation based on the perturbative calculation in presence of 
interaction and disorder has been proposed and it seems to give a 
quantitative estimate closer to the experimental results, but still it 
is less than the measured currents by an order of magnitude, and the 
interaction parameter used in the theory is not well understood 
physically. Most of these theoretical results have been obtained based
on a tight-binding framework within the {\em nearest-neighbor hopping} 
(NNH) approximation. This is an important approximation and it has been
shown that within the NNH model electronic correlation provides a 
small enhancement of current amplitude in disordered materials i.e., a
weak delocalizing effect is observed in presence of electron-electron 
(e-e) interaction. As an attempt in the present work we modify the 
traditional NNH model by incorporating the effects of higher order 
hopping integrals, at least second-neighbor hopping (SNH), in addition 
to the NNH integral. It is also quite physical since electrons have some 
finite probabilities to hop from one site to other sites apart from 
nearest-neighbor with reduced strengths. We will show that the inclusion 
of higher order hopping integrals gives significant enhancement of current 
amplitude and it reaches quite closer to the current amplitude of ordered 
systems. 

The other important controversy comes for the determination of the sign 
of low-field currents and still it is an unresolved issue between 
theoretical and experimental results. In an experiment on persistent 
current Levy {\em et al.}~\cite{levy} have shown diamagnetic nature for 
the measured currents at low-field limit. While, in other experiment 
Chandrasekhar {\em et al.}~\cite{chand} have obtained paramagnetic response 
near zero field limit. Jariwala {\em et al.}~\cite{jari} have predicted 
diamagnetic persistent current in their experiment and similar diamagnetic 
response in the vicinity of zero field limit were also supported in an 
experiment done by Deblock~\cite{deb} {\em et al.} on Ag rings. Yu and 
Fowler~\cite{yu} have shown both diamagnetic and paramagnetic responses 
in mesoscopic Hubbard rings. Though in a theoretical work Cheung {\em et 
al.}~\cite{cheu2} have predicted that the direction of current is random 
depending on the total number of electrons in the system and the specific 
realization of the random potentials. Hence, prediction of the sign of 
low-field currents is still an open challenge and further studies on 
persistent current in mesoscopic systems are needed to remove the existing 
controversies.

In the present paper we address the behavior of persistent current in an
interacting mesoscopic ring with finite width threaded by an Aharonov-Bohm 
flux $\phi$. A simple tight-binding Hamiltonian is used to illustrate the 
system and all the calculations are performed within a mean field approach. 
Using a generalized Hartree-Fock (HF) approximation~\cite{kato,kam}, 
we compute numerically persistent current ($I$) as functions of AB flux 
$\phi$, total number of electrons $N_e$, e-e interaction strength $U$,
second-neighbor hopping integral, disorder strength $W$ and system size $N$.
{\em The main motivation of the present work is to illustrate the effects of 
higher order hopping integrals on the enhancement of persistent current
in disordered mesoscopic cylinders.} Our results can be utilized to study 
magnetic response in any interacting mesoscopic system. 

In what follows, we present the results. In Section II, we describe the 
geometric model and generalized Hartree-Fock theory to study magnetic 
response in the model quantum system. Section III contains the numerical 
results. Finally, in Section IV we draw our conclusions.

\section{Model and synopsis of the theoretical formulation}

\begin{figure}[ht]
{\centering \resizebox*{5cm}{3cm}
{\includegraphics{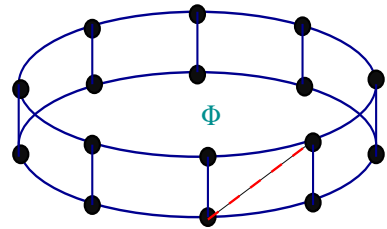}}\par}
\caption{(Color online). Schematic view of a $1$D mesoscopic cylinder 
penetrated by a magnetic flux $\phi$. The red dashed line corresponds to 
the second-neighbor hopping integral and the filled black circles represent 
the positions of the atomic sites. A persistent current $I$ is established 
in the cylinder.}
\label{cylinder}
\end{figure}

Let us start by referring to Fig.~\ref{cylinder}, where a small metallic 
cylinder is threaded by a magnetic flux $\phi$. The filled black circles 
correspond to the positions of the atomic sites in the cylinder. To predict
the size of a cylinder we use two parameters $N$ and $M$, where the $1$st
one ($N$) represents total number of atomic sites in each circular ring 
and the other one ($M$) gives total number of identical circular rings. 
For the description of our model quantum system we use a tight-binding 
(T-B) framework and in order to incorporate the effect of higher order
hopping integrals to the Hamiltonian here we consider second-neighbor 
hopping (SNH) (shown by the red dashed line in Fig.~\ref{cylinder}) in 
addition to the nearest-neighbor hopping (NNH) of electrons. Considering 
both NNH and SNH integrals the T-B Hamiltonian for the cylindrical system 
in Wannier basis looks in the form,
\begin{eqnarray}
H_{\mbox{c}} & = &\sum_{i,j,\sigma}\epsilon_{i,j,\sigma} 
c_{i,j,\sigma}^{\dagger} c_{i,j,\sigma} + \sum_{i,j,\sigma} 
t_l \left[e^{i\theta_l} c_{i,j,\sigma}^{\dagger} c_{i,j+1,\sigma}
\right. \nonumber \\ 
& + & \left. h.c. \right] + \sum_{i,j,\sigma} t_d \left[e^{i\theta_d} 
c_{i,j,\sigma}^{\dagger} c_{i+1,j+1,\sigma} + h.c. \right] \nonumber \\
& + & \sum_{ij} U c_{i,j,\uparrow}^{\dagger}c_{i,j,\uparrow} 
c_{i,j,\downarrow}^{\dagger} c_{i,j,\downarrow}
\label{equ1}
\end{eqnarray}
where, ($i,j$) represent the co-ordinate of a lattice site. The index
$i$ runs from $1$ to $M$, while the integer $j$ goes from $1$ to $N$. 
$\epsilon_{i,j,\sigma}$ is the on-site energy of an electron at the site 
($i,j$) of spin $\sigma$ ($\uparrow,\downarrow$). $t_l$ and $t_d$ are the 
NNH and SNH integrals, respectively. Due to the presence of magnetic flux 
$\phi$ (measured in unit of the elementary flux quantum $\phi_0=ch/e$), a 
phase factor $\theta_l=2\pi\phi/N$ appears in the Hamiltonian when an 
electron hops longitudinally from one site to its neighboring site, 
and accordingly, a negative sign comes when the electron hops in the 
reverse direction. $\theta_d$ is the associated phase factor for the 
diagonal motion of an electron between two neighboring concentric rings.
No phase factor appears when an electron moves along the vertical direction 
which is set by proper choice of the gauge for the vector potential 
$\vec{A}$ associated with the magnetic field $\vec{B}$, and this choice 
makes the phase factors ($\theta_l$, $\theta_d$) identical to each other 
for the longitudinal and diagonal motions. Since the magnetic 
field corresponding to the AB flux $\phi$ does not penetrate anywhere of 
the surface of the cylinder, we ignore Zeeman term in the above 
tight-binding Hamiltonian (Eq.~\ref{equ1}). $c_{i,j,\sigma}^{\dagger}$ 
and $c_{i,j,\sigma}$ are the creation and annihilation operators, 
respectively, of an electron at the site ($i,j$) with spin $\sigma$. 
$U$ is the on-site Hubbard interaction term. 

\subsection{Decoupling of the interacting Hamiltonian}

To get the energy eigenvalues of the interacting model quantum system 
described by the above tight-binding Hamiltonian given in Eq.~\ref{equ1}, 
first we decouple the interacting Hamiltonian using generalized Hartree-Fock 
approach, the so-called mean field approximation. In this procedure, the 
full Hamiltonian is completely decoupled into two parts. One is associated 
with the up-spin electrons, while the other is related to the down-spin 
electrons with their modified site energies. For up and down spin 
Hamiltonians, the modified site energies are expressed in the form,
\begin{equation}
\epsilon_{i,j,\uparrow}^{\prime}=\epsilon_{i,j,\uparrow} + U \langle 
n_{i,j,\downarrow} \rangle
\label{equ2}
\end{equation}
\begin{equation}
\epsilon_{i,j,\downarrow}^{\prime}=\epsilon_{i,j,\downarrow} + U \langle 
n_{i,j,\uparrow} \rangle
\label{equ3}
\end{equation}
where, $n_{i,j,\sigma}=c_{i,j,\sigma}^{\dagger} c_{i,j,\sigma}$ is the 
number operator. With these site energies, the full Hamiltonian 
(Eq.~\ref{equ1}) can be written in the decoupled form as,
\begin{eqnarray}
H_{\mbox{c}} &=&\sum_{i,j} \epsilon_{i,j,\uparrow}^{\prime} n_{i,j,\uparrow} 
+ \sum_{i,j} t_l \left[e^{i\theta_l} 
c_{i,j,\uparrow}^{\dagger} c_{i,j+1,\uparrow} + h.c.\right] \nonumber \\
& + & \sum_{i,j} t_d \left[e^{i\theta_d} 
c_{i,j,\uparrow}^{\dagger} c_{i+1,j+1,\uparrow} + h.c.\right] 
\nonumber \\
& + & \sum_{i,j} \epsilon_{i,j,\downarrow}^{\prime} n_{i,j,\downarrow} 
+ \sum_{i,j} t_l \left[e^{i\theta_l} c_{i,j,\downarrow}^{\dagger} 
c_{i,j+1,\downarrow} + h.c. \right] \nonumber \\ 
& + & \sum_{i,j} t_d \left[e^{i\theta_d} 
c_{i,j,\downarrow}^{\dagger} c_{i+1,j+1,\downarrow} + h.c.\right] 
\nonumber \\
& - & \sum_{i,j} U \langle n_{i,j,\uparrow} \rangle \langle 
n_{i,j,\downarrow} \rangle \nonumber \\
&=& H_{\uparrow}+H_{\downarrow}-\sum_{i,j} U \langle n_{i,j,\uparrow} 
\rangle \langle n_{i,j,\downarrow} \rangle
\label{equ4} 
\end{eqnarray}\\
where, $H_{\uparrow}$ and $H_{\downarrow}$ correspond to the effective
tight-binding Hamiltonians for the up and down spin electrons, respectively.
The last term is a constant term which provides an energy shift in the 
total energy. 

\subsection{Self consistent procedure}

With these decoupled Hamiltonians ($H_{\uparrow}$ and $H_{\downarrow}$) 
of up and down spin electrons, now we start our self consistent procedure 
considering initial guess values of $\langle n_{i,j,\uparrow} \rangle$ and 
$\langle n_{i,j,\downarrow} \rangle$. For these initial set of values of 
$\langle n_{i,j,\uparrow} \rangle$ and $\langle n_{i,j,\downarrow} \rangle$, 
we numerically diagonalize the up and down spin Hamiltonians. Then we 
calculate a new set of values of $\langle n_{i,j,\uparrow} \rangle$ and 
$\langle n_{i,j,\downarrow} \rangle$. These steps are repeated until a 
self consistent solution is achieved.

\subsection{Calculation of ground state energy}

After achieving the self consistent solution, the ground state energy 
$E_0$ for a particular filling at absolute zero temperature ($T=0$K)
can be determined by taking the sum of individual states up to Fermi 
energy ($E_F$) for both up and down spins. Thus, we can write the final
form of ground state energy as,
\begin{equation}
E_0=\sum_p E_{p,\uparrow} + \sum_p E_{p,\downarrow}- \sum_{i,j} U 
\langle n_{i,j,\uparrow} \rangle \langle n_{i,j,\downarrow} \rangle
\label{equ5} 
\end{equation} 
where, the index $p$ runs for the states up to the Fermi level. 
$E_{p,\uparrow}$ ($E_{p,\downarrow}$) is the single particle energy 
eigenvalue for $p$-th eigenstate obtained by diagonalizing the Hamiltonian 
$H_{\uparrow}$ ($H_{\downarrow}$).

\subsection{Calculation of persistent current} 

At absolute zero temperature, total persistent current of the system
is obtained from the expression,
\begin{equation}
I(\phi)=-c\frac{\partial E_0(\phi)}{\partial \phi}
\label{equ6}
\end{equation}
where, $E_0(\phi)$ is the ground state energy for a particular filling.

In the present work we perform all the essential features of persistent
current at absolute zero temperature and use the units where $c=h=e=1$.
Throughout our numerical calculations we set the nearest-neighbor hopping 
strength $t_l=-1$ and fix $M=2$ i.e., cylinders with two identical rings. 
Energy scale is measured in unit of $t_l$.

\section{Numerical results and discussion}

Following the above theoretical prescription now we start to analyze our
numerical results. We describe the results in three different parts. In 
the first part, we consider perfect cylinders with only nearest-neighbor 
hopping integral. In the second part, disordered cylinders described with 
only NNH integral are considered. Finally, in the third part we discuss 
the effect of second-neighbor hopping (SNH) integral on the enhancement 
of persistent current in disordered cylinders.

\subsection{Perfect cylinders with NNH integral}

For perfect cylinders we choose $\epsilon_{i,j,\uparrow}=
\epsilon_{i,j,\downarrow}=0$ for all ($i,j$). Since here we consider the 
cylinders described with NNH integral only, the second-neighbor hopping 
strength $t_d$ is fixed to zero. 

\subsubsection{Energy-flux characteristics}

\begin{figure}[ht]
{\centering \resizebox*{7.5cm}{7cm}
{\includegraphics{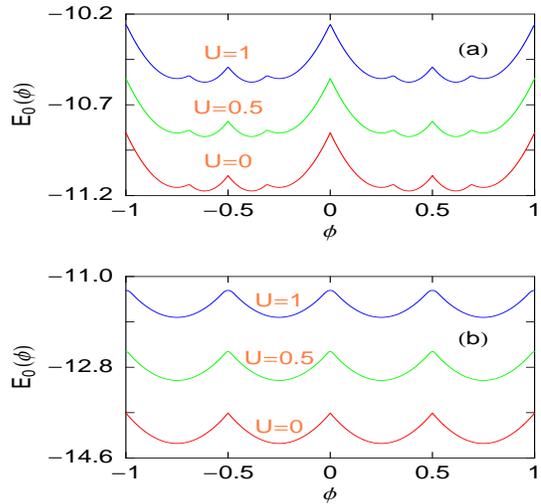}}\par}
\caption{(Color online). Ground state energy levels as a function of 
flux $\phi$ for some perfect cylinders with $N=5$ and $M=2$. The red, 
green and blue curves correspond to $U=0$, $0.5$ and $1$, respectively. 
(a) Quarter-filled case and (b) Half-filled case.}
\label{cylinderenergy1}
\end{figure}

As illustrative examples, in Fig.~\ref{cylinderenergy1} we show the 
variation of ground state energy levels as a function of magnetic flux 
$\phi$ for some typical mesoscopic cylinders where $N$ is fixed at $5$
(odd $N$). In (a) the results are given for the quarterly-filled ($N_e=5$) 
cylinders, while in (b) the curves correspond to the results for the 
half-filled ($N_e=10$) cylinders.
The red, green and blue lines represent the ground state energy levels for 
$U=0$, $0.5$ and $1$, respectively. It is observed that the ground state
energy shows oscillatory behavior as a function of $\phi$ and the energy
increases as the electronic correlation strength $U$ gets increased. Most 
significantly we see that the ground state energy levels provide two 
\begin{figure}[ht]
{\centering \resizebox*{7.5cm}{7cm}
{\includegraphics{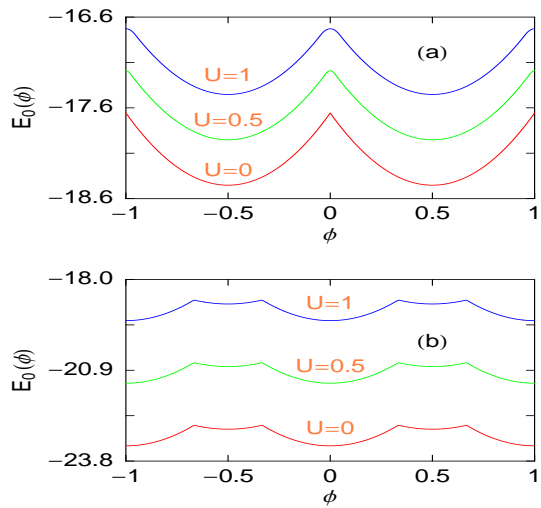}}\par}
\caption{(Color online). Ground state energy levels as a function of 
flux $\phi$ for some perfect cylinders considering $N=8$ and $M=2$. The 
red, green and blue curves correspond to $U=0$, $0.5$ and $1$, respectively. 
(a) Quarter-filled case and (b) Half-filled case.}
\label{cylinderenergy2}
\end{figure}
different types of periodicities depending on the electron filling. At 
quarter-filling, ground state energy level gives $\phi_0$ ($=1$, since 
$c=e=h=1$ in our chosen unit system) flux-quantum periodicity. On the 
other hand, at half-filling it shows $\phi_0/2$ flux-quantum periodicity.
The situation becomes quite different when the total number of atomic sites 
$N$ in individual rings is even. For our illustrative purposes in 
Fig.~\ref{cylinderenergy2} we plot the lowest energy levels as a function 
of $\phi$ for some typical mesoscopic cylinders considering $N=8$ (even $N$). 
The curves of different colors correspond to the identical meaning as in 
Fig.~\ref{cylinderenergy1}. From the spectra given in 
Figs.~\ref{cylinderenergy2}(a) (quarter-filled case) and (b) (half-filled
case) it is clearly observed that the ground state energy levels vary 
periodically with AB flux $\phi$ exhibiting only $\phi_0$ flux-quantum 
periodicity. Thus it can be emphasized that the appearance of half 
flux-quantum periodicity strongly depends on the electron filling as well 
as on the oddness and evenness of the total number of atomic sites $N$ in 
individual rings. Only for the half-filled cylinders with odd $N$, the 
lowest energy level gets $\phi_0/2$ periodicity with flux $\phi$. Now it
is important to note that this half flux-quantum periodicity does not 
depend on the width ($M$) of the cylinder and also it is independent of 
the Hubbard correlation strength $U$. Hence, depending on the system size 
and filling of electrons variable 
periodicities are observed in the variation of lowest energy level. It 
may provide an important signature in studying magnetic response in 
nano-scale loop geometries.

\subsubsection{Current-flux characteristics}

\begin{figure}[ht]
{\centering \resizebox*{7.5cm}{7cm}
{\includegraphics{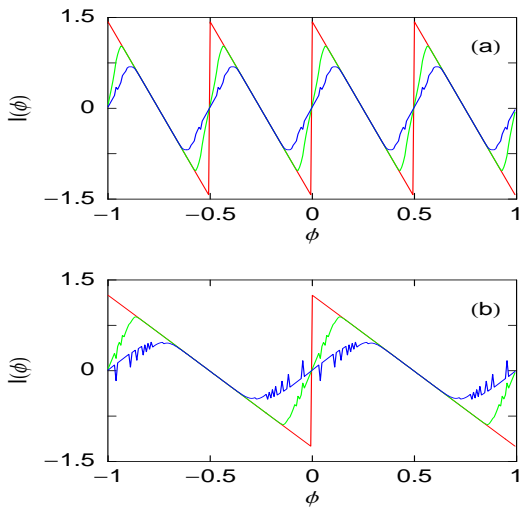}}\par}
\caption{(Color online). Persistent current as a function of flux $\phi$ 
for some ordered mesoscopic cylinders considering $M=2$. (a) Half-filled
case with $N=15$. The red, green and blue curves correspond to $U=0$, $1.5$ 
and $2$, respectively. (b) Quarter-filled case with $N=20$. The red, green 
and blue curves correspond to $U=0$, $2$ and $3$, respectively.}
\label{cylindercurr}
\end{figure}

In Fig.~\ref{cylindercurr} we display the current-flux characteristics
for some impurity free mesoscopic cylinders considering $M=2$. In (a) the 
results are given for the half-filled case where we set $N=15$. The red 
line corresponds to the current for the non-interacting ($U=0$) case, 
while the green and blue lines represent the currents when $U=1.5$ and 
$2$, respectively. From the curves we notice that the current amplitude
gradually decreases with the increase of electronic correlation strength
$U$. The reason is that at half-filling each site is occupied by at 
least one electron of up spin or down spin, and the placing of a second
electron of opposite spin needs more energy due to the repulsive effect
of $U$. Thus conduction becomes difficult as it requires more energy when 
an electron hops from its own site and situates at the neighboring site. 
Now both for the non-interacting and interacting cases, current shows
half flux-quantum periodicity as a function a $\phi$ obeying the 
energy-flux characteristics since here we choose odd $N$ ($N=15$). The 
behavior of the persistent currents for even $N$ is shown in (b) where
we set $N=20$. The currents are drawn for the quarter-filled case i.e., 
$N_e=20$, where the red, green and blue curves correspond to $U=0$, 
$2$ and $3$, respectively. The reduction of current amplitude with the
increase of Hubbard interaction strength is also observed for this
quarter-filled case, similar to the case of half-filled as described 
earlier. But the point is that at quarter-filling, the reduction of 
current amplitude is much smaller compared to the half-filled situation.
This is quite obvious in the sense that at less than half-filling `empty' 
lattice sites are available where electrons can hop easily without any cost
of extra energy and the conduction becomes much easier than the half-filled 
situation. In this quarter-filled case, persistent currents provide only 
$\phi_0$ flux-quantum periodicity following the $E$-$\phi$ diagram. From 
these current-flux characteristics it can be concluded that for {\em 
`ordered' cylinders current amplitude always decreases with the enhancement 
in Hubbard correlation strength $U$.} 

\subsection{Disordered cylinders with NNH integral}

In order to describe the effect of impurities on electron transport now 
we focus our attention on the results of some typical disordered cylinders 
described with NNH integral. Here we consider the diagonal disordered
cylinders i.e., impurities are introduced only at the site energies 
without disturbing the hopping integrals. The site energies in each 
concentric ring are chosen from a correlated distribution function 
which looks in the form,
\begin{equation}
\epsilon_{j,\uparrow}=\epsilon_{j,\downarrow}=W \cos\left(j \lambda 
\pi\right)
\label{equ7}
\end{equation} 
where, $W$ is the impurity strength. $\lambda$ is an irrational number 
and we choose $\lambda=(1+\sqrt{5})/2$, for the sake of our illustration. 
Setting $\lambda=0$, we get back the pure system with uniform site energy 
$W$. Now, instead of considering site energies from a correlated
distribution function, as mentioned above in Eq.~\ref{equ7}, we can also 
take them randomly from a ``Box" distribution function of width $W$.
But in the later case we have to take the average over a large number of
disordered configurations (from the stand point of statistical average)
and since it is really a difficult task in the aspect of numerical
computation we select the other option. Not only that in the averaging
process several mesoscopic phenomena may disappear. Therefore, the 
averaging process is an important issue in low-dimensional systems.

\begin{figure}[ht]
{\centering \resizebox*{7.5cm}{7cm}
{\includegraphics{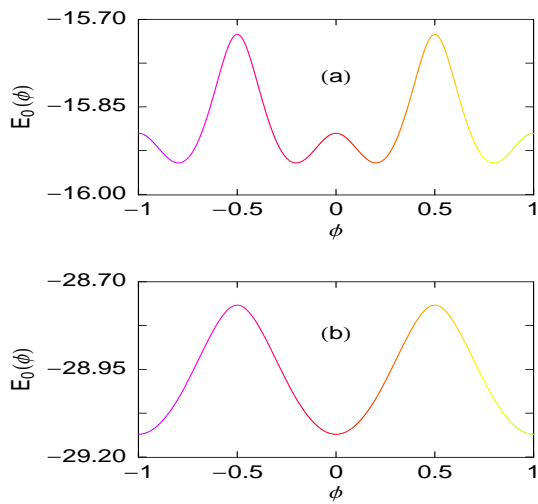}}\par}
\caption{(Color online). Ground state energy level as a function of 
flux $\phi$ for half-filled disordered mesoscopic cylinders ($M=2$) 
considering $U=1$ and $W=2$. (a) $N=5$ and (b) $N=8$.}
\label{cylinderenergy3}
\end{figure}

In presence of disorder, energy levels get modified significantly. For our
illustrative purposes in Fig.~\ref{cylinderenergy3} we plot ground state 
energy levels as a function of magnetic flux $\phi$ for some disordered 
mesoscopic cylinders when they are half-filled. The Hubbard interaction 
strength $U$ is set at $1$ and the impurity strength $W$ is fixed to $2$. 
In (a) the ground state energy level is shown for a cylinder with $N=5$ 
(odd), while in (b) it is presented for a cylinder taking $N=8$ (even). 
Quite interestingly we see that for the cylinder with odd $N$, the half 
flux-quantum periodicity of the lowest energy level disappears in the
presence of impurity and it provides conventional $\phi_0$ periodicity. 
Hence, for cylinders with odd $N$, $\phi_0/2$ flux-quantum periodicity 
will be observed only when they are free from any impurity. For the 
disordered cylinder with even $N$ ($N=8$), the lowest energy level 
as usual provides $\phi_0$ periodicity similar to the impurity free 
cylinders containing even $N$. Apart from this periodic nature, impurities 
play another significant role in the determination of the slope of the 
energy levels. The slope of the lowest energy level decreases significantly
compared to the perfect case, and therefore, a prominent change in current
amplitude also takes place. 

\begin{figure}[ht]
{\centering \resizebox*{7.75cm}{4.3cm}
{\includegraphics{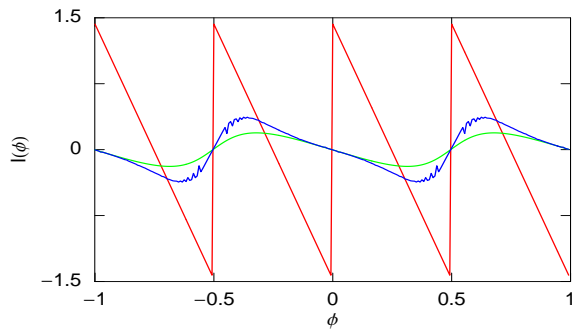}}\par}
\caption{(Color online). Persistent current as a function of flux $\phi$ 
for a half-filled mesoscopic cylinder considering $N=15$ and $M=2$. The 
red line corresponds to the ordered case when $U=0$, whereas the green
and blue lines correspond to the disordered case ($W=2$) when $U=0$ and 
$1.5$, respectively.}
\label{cylindercurr1}
\end{figure}

To justify the above facts, in Fig.~\ref{cylindercurr1} we present the 
variations of persistent currents with AB flux $\phi$ for a half-filled
mesoscopic cylinder, described in the framework of NNH model, considering 
$N=15$ and $M=2$. The red curve represents the current for the ordered 
($W=0$) non-interacting ($U=0$) cylinder. It shows saw-tooth like nature 
with flux $\phi$ providing $\phi_0/2$ flux-quantum periodicity. The 
situation becomes completely different when impurities are introduced in 
the cylinder as seen by the other two curves. The green curve represents 
the current for the case only when impurities are considered but the 
effect of electronic correlation is not taken into account. It shows a
continuous like nature with $\phi_0$ flux-quantum periodicity. The most
important observation is that the current amplitude gets reduced enormously, 
even an order of magnitude, compared to the perfect cylinder.
This is due to the localization of the energy eigenstates in the presence 
of impurity, which is the so-called Anderson localization. Hence, a large 
difference exists in the current amplitudes of an ordered and disordered 
non-interacting cylinders and it was the main controversial issue among 
the theoretical and experimental predictions. Experimental verifications 
suggest that the measured current amplitude is quite comparable to the
theoretical current amplitude obtained in a perfect system. To remove this 
controversy, as a first attempt, we include the effect of e-e correlation
in the disordered cylinder described by the NNH model. The result is shown 
by the blue curve where $U$ is fixed at $1.5$. It is observed that the 
current amplitude gets increased compared to the non-interacting 
disordered cylinder, though the increment is too small. Not only that 
the enhancement can take place only for small values of $U$, while for 
large enough $U$ the current amplitude rather decreases. This phenomenon 
can be implemented as follows. 
For the non-interacting disordered cylinder the probability of getting two 
opposite spin electrons becomes higher at the atomic sites where the site 
energies are lower than the other sites since the electrons get pinned at 
the lower site energies to minimize the ground state energy, and this
pinning of electrons becomes increased with the rise of impurity strength 
$W$. As a result the mobility of electrons and hence the current amplitude 
gets reduced with the increase of impurity strength $W$. Now, if we 
introduce electronic correlation in the system then it tries to depin 
two opposite spin electrons those are situated together due to the Coulomb 
repulsion. Therefore, the electronic mobility is enhanced which provides
larger current amplitude. But, for large enough interaction strength, no 
electron can able to hop from one site to other at the half-filling 
since then each site is occupied either by an up or down spin electron
which does not allow other electron of opposite spin due to the repulsive
term $U$. Accordingly, the current amplitude gradually decreases with $U$.
On the other hand, at less than half-filling though there is some finite 
probability to hop an electron from one site to the other available 
`empty' site but still it is very small. So, in brief, we can say that 
within the nearest-neighbor hopping (NNH) approximation electron-electron 
interaction does not provide any significant contribution to enhance 
the current amplitude, and hence the controversy regarding the current 
amplitude still persists.

\subsection{Disordered cylinders with NNH and SNH integrals}

To overcome the existing situation regarding the current amplitude, 
in this sub-section, finally we make an attempt by incorporating the 
effect of second-neighbor hopping (SNH) integral in addition to the 
nearest-neighbor hopping (NNH) integral. 

\begin{figure}[ht]
{\centering \resizebox*{7.75cm}{4.3cm}
{\includegraphics{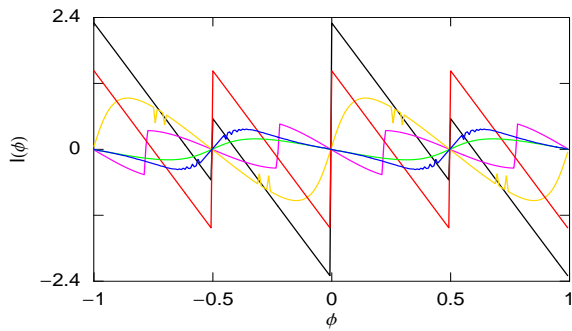}}\par}
\caption{(Color online). Persistent current as a function of flux $\phi$ 
for a half-filled mesoscopic cylinder taking $N=15$ and $M=2$ in the
presence of NNH and SNH integrals. The black line corresponds to the 
ordered case when $U=0$, whereas the magenta and gold lines correspond 
to the disordered case ($W=2$) when $U=0$ and $1.5$, respectively. Here
SNH integral is fixed at $-0.6$. The currents shown by the red, green 
and blue lines for the ring described with NNH model (identical to 
Fig.~\ref{cylindercurr1}) are re-plotted to judge the effect of SNH 
integral over NNH model much clearly.}
\label{cylindercurr2}
\end{figure}

A significant change in current amplitude takes place when we include 
the contribution of second-neighbor hopping (SNH) integral in addition to 
the NNH integral. As representative examples, in Fig.~\ref{cylindercurr2} 
we plot the current-flux characteristics for a half-filled mesoscopic 
cylinder considering $N=15$ and $M=2$. The black, 
magenta and gold lines correspond to the results in the presence of 
SNH integral, while the other three colored curves (red, green and 
blue) represent the currents in the absence of SNH integral. Here 
we choose $t_d=-0.6$. The black curve refers to the persistent current 
for the perfect ($W=0$) non-interacting ($U=0$) cylinder and it achieves 
much higher amplitude compared to the NNH model (red curve). This 
additional contribution comes from the SNH integral since it allows 
electrons to hop further. In addition it is also noticed that the current 
varies periodically with $\phi$ providing $\phi_0$ flux-quantum periodicity, 
instead of $\phi_0/2$ as in the case of NNH integral model (red curve). 
Thus, it can be emphasized that $\phi_0/2$ periodicity will be observed 
only when the cylinder is (a) free from impurity, (b) half-filled, (c) 
made with odd $N$, and (d) described by the nearest-neighbor hopping model.
The main focus of this sub-section is to interpret the combined effect
of SNH integral and electron-electron correlation on the enhancement of
persistent current amplitude in disordered cylinder. To do this first we 
narrate the effect of SNH integral in disordered non-interacting cylinder. 
The nature of the current for this particular case is shown by the magenta 
curve of Fig.~\ref{cylindercurr2}. It shows that the current amplitude 
gets reduced compared to the perfect case (black line), which is expected, 
but the reduction of the current amplitude is quite small than the NNH 
integral model. This is due the fact that the SNH integral tries to 
delocalize the electronic 
\begin{figure}[ht]
{\centering \resizebox*{7.75cm}{4.3cm}
{\includegraphics{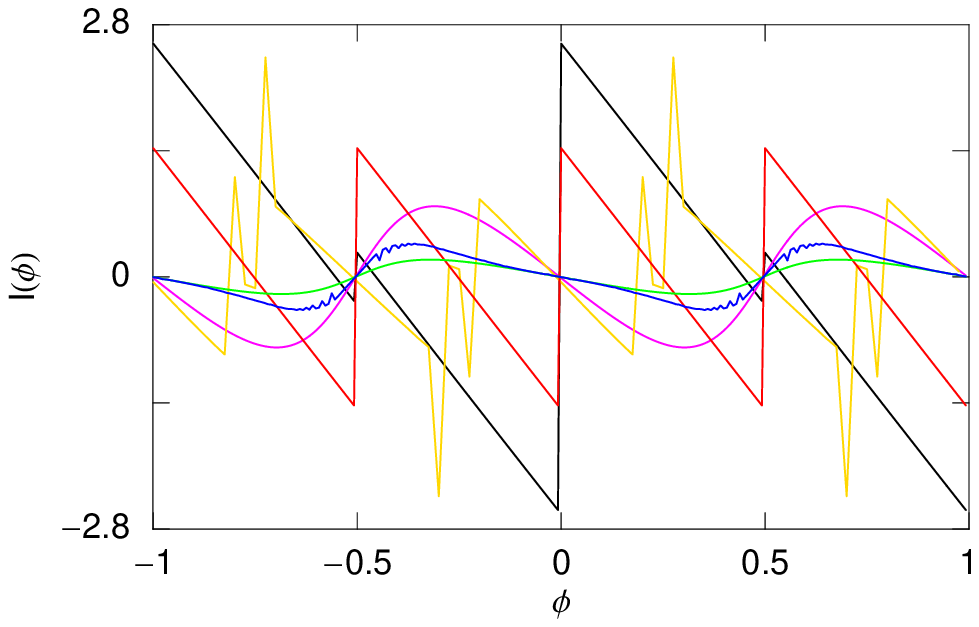}}\par}
\caption{(Color online). Persistent current as a function of flux $\phi$ 
for a half-filled mesoscopic cylinder taking $N=15$ and $M=2$ in the
presence of NNH and SNH integrals. The black line corresponds to the 
ordered case when $U=0$, whereas the magenta and gold lines correspond 
to the disordered case ($W=2$) when $U=0$ and $1.5$, respectively. Here
SNH integral is fixed at $-0.8$. The currents shown by the red, green 
and blue lines for the ring described with NNH model (identical to 
Fig.~\ref{cylindercurr1}) are re-plotted to judge the effect of SNH 
integral over NNH model much clearly.}
\label{cylindercurr3}
\end{figure}
states, and therefore, the mobility of the electrons is enriched. The 
situation becomes more interesting when we include the effect of Hubbard 
interaction. The behavior of the current in the presence of interaction 
is plotted by the gold curve of Fig.~\ref{cylindercurr2} where we fix 
$U=1.5$. Very interestingly we see that the current amplitude is enhanced 
significantly and quite comparable to that of the perfect cylinder.

For better clarity of the results discussed above, in 
Fig.~\ref{cylindercurr3} we also present the similar feature of persistent
current for other hopping strength of SNH integral. Here we set $t_d=-0.8$. 
From these curves we see that the current amplitude gets enhanced more as 
we increase the SNH strength.

Thus, it can be predicted that the presence of SNH integral 
and Hubbard interaction can provide a persistent current which may be 
comparable to the measured current amplitudes. In this presentation we 
consider the effect of only SNH integral as a higher order hopping integral 
in addition to the NNH model, and, illustrate how such a higher order 
hopping integral leads an important role on the enhancement of
current amplitude in presence of Hubbard correlation for disordered 
cylinders. Instead of considering only the SNH integral we can also take 
the contributions from all possible higher order hopping integrals with 
reduced hopping strengths. Since the strengths of other higher order 
hopping integrals are too small, the contributions from these factors 
are reasonably small and they will not provide any significant change in 
the current amplitude. Finally, we can say that further studies are 
needed by incorporating all these factors. 

\section{Closing remarks}

To summarize, in the present work we have addressed the behavior of
persistent current in an interacting mesoscopic cylinder threaded by 
an Aharonov-Bohm flux $\phi$. We have adopted a tight-binding Hamiltonian
to describe the model quantum system and all the numerical calculations 
have been done within a mean field approximation. Using the generalized 
Hartree-Fock (HF) approximation, we have computed persistent current as
functions of SNH integral, impurity strength $W$, AB flux $\phi$, electron 
filling $N_e/N$ and system size $N$. Our numerical results have provided
several interesting features and the present study may be helpful in 
understanding magnetic response in nano-scale loop geometries.

The essential features observed from our analysis are as follows. \\
(i) In the determination of the lowest energy level we see that the 
energy level varies periodically with $\phi$ exhibiting both $\phi_0/2$
and $\phi_0$ flux-quantum periodicities depending on the choices of
the parameters describing the tight-binding Hamiltonian. \\
(ii) In the NNH model, current amplitude gets significantly reduced when
impurities are introduced in the system. With the inclusion of Hubbard
interaction ($U$), the current amplitude can be enhanced, though the 
enhancement becomes too small compared to the experimental verifications. \\
(iii) A significant change in the current amplitude takes place when the
effect of SNH integral is taken into account. The combined effect of SNH
and Hubbard interaction can provide the current which is quite comparable
to the experimental realizations. So in short we can say that the 
conventional NNH model can be modified by incorporating the higher order 
hopping integrals.

Throughout the analysis we have kept the width of the cylinders at a fixed
value ($M=2$), for the sake of our illustration. All these results are also
valid for cylinders of larger widths. Here we have considered several 
important approximations by ignoring the effects of temperature, 
electron-phonon interaction, etc. Due to these factors, any scattering 
process that appears in the cylinder would have influence on electronic 
phases. At the end, we would like to say that we need further study in 
such systems by incorporating all these effects.

\vskip 0.3in
\noindent
{\bf\small ACKNOWLEDGMENTS}
\vskip 0.2in
\noindent
I acknowledge with deep sense of gratitude the illuminating comments
and suggestions I have received from Prof. S. N. Karmakar during the 
calculations.

\end{document}